\documentclass[
  journal=pasa,
  manuscript=research-paper, 
  year=2025,
  volume=YY,
]{cup-journal}

\usepackage{amsmath}
\usepackage{amssymb,microtype,siunitx,booktabs}
\newcommand\ion[2]{\text{#1\,\textsc{\lowercase{#2}}}}
\newcommand{\HI}{\ion{H}{I}}

\usepackage[breaklinks,colorlinks=true,citecolor=blue,linkcolor=Purple,urlcolor=Violet,bookmarks=true]{hyperref}

\title{ Orientation of galaxy spins relative to filaments of the large-scale structure of the Universe}
\author{A.~V.~Antipova}
\affiliation{Special Astrophysical Observatory, Russian Academy of Sciences, Nizhnii Arkhyz 369167, Russia} 
\alsoaffiliation{Central (Pulkovo) Astronomical Observatory, Russian Academy of Sciences, Pulkovskoye chaussee 65/1, St. Petersburg 196140, Russia}
\email[A. Antipova]{antal@sao.ru}

\author{D.~I.~Makarov}
\affiliation{Special Astrophysical Observatory, Russian Academy of Sciences, Nizhnii Arkhyz 369167, Russia} 
\alsoaffiliation{Central (Pulkovo) Astronomical Observatory, Russian Academy of Sciences, Pulkovskoye chaussee 65/1, St. Petersburg 196140, Russia}

\author{N.~I.~Libeskind}
\affiliation{Leibniz-Institut fur Astrophysik Potsdam (AIP), An der Sternwarte 16, D-14482 Potsdam, Germany}

\author{E.~Tempel}
\affiliation{Tartu Observatory, University of Tartu, Observatooriumi 1, 61602 T\~oravere, Estonia}

\doi{10.1017/pasa.2025.10041}

\received {24 November 2024}
\revised  {22 April 2025}
\accepted {1 May 2025}
\published{}


\begin{document}
\begin{abstract}
The theory of galaxy formation posits a clear correlation between the spin of galaxies and the orientation of the elements of the large-scale structure of the Universe, particularly cosmic filaments. 
A substantial number of observational and modelling studies have been undertaken with the aim of identifying the dependence of spin orientation on the components of the large-scale structure. 
However, the findings of these studies remain contradictory.
In this paper, we present an analysis of the orientation of the spins of 2,861 galaxies with respect to the filaments of the large-scale structure of the Universe. 
All galaxies in our sample have an inclination to the line of sight greater than 85 degrees, enabling an unambiguous determination of the spin axis direction in space. 
We investigate the alignment of galaxy spin axes relative to cosmic web filaments as a function of various properties for galaxies. 
Our results reveal a statistically significant tendency for the galaxy spin axes to align along the filament axes of the large-scale structure.
\end{abstract}

\section{Introduction}
\label{sec:Intro}

The distribution of galaxies in the Universe forms a large-scale structure consisting of voids, filaments, sheets, and clusters.
Most galaxies reside in filaments that lie between voids and connect galaxy clusters. 
The theory of galaxy formation posits that the angular momentum (spin) of protogalaxies arises under the influence of tidal forces exerted by elements of the large-scale structure of the Universe \citep{1979MNRAS.186..133E,1969ApJ...155..393P,2002MNRAS.332..325P,1984ApJ...286...38W}. 
According to this theory, the axis of rotation of a protogalaxy should correlate with the principal axes of the local tidal shear tensors~\citep{1992ApJ...401..441D,2000ApJ...532L...5L}.
Consequently, if the original angular momentum of a protogalaxy is preserved, a clear correlation between the galaxy spins and the elements of the large-scale structure of the Universe should be observed.

A substantial amount of works, both observational and modelling-based, has been undertaken to search for the dependence of the spin orientation relative to elements of the large-scale structure.
However, the results remain contradictory.

For instance, when studying the orientation of the spins of 100 galaxies relative to filaments in a hydrodynamic cosmological simulation, it was found that the most massive disk galaxies at all redshifts tend to be oriented along the filament~\citep{2010MNRAS.405..274H}. Conversely, based on N-body modelling, \citet{2007ApJ...655L...5A,2012MNRAS.427.3320C,2012MNRAS.421L.137L,2013ApJ...762...72T} found that massive halos exhibit a perpendicular spin orientation relative to the filaments, while the spins of low-mass halos are aligned along the filament axes. The mass of the halo at which the spin orientation transition occurs is approximately $(5 \pm 1) \times 10^{12}$~M${\odot}$ \citep{2012MNRAS.427.3320C}. However, subsequent studies have shown that the mass at which the spin orientation transition occurs is not universal and depends on the properties of the large-scale structure itself \citep{2013MNRAS.428.2489L}. The dependence of spin orientation on mass was identified by \citet{2018ApJ...866..138W}, where the transition mass is approximately $10^{9.4}$~h$^{-1}$~M${\odot}$; a correlation with the colour index was also discovered. However, \citet{2019MNRAS.487.1607G} reported a perpendicular orientation of spins relative to filaments, regardless of mass. Additionally, \citet{2019MNRAS.483.3227K} found a dependence on the \HI{} mass: galaxies with high \HI{} exhibit a tendency to have a parallel spin orientation, while galaxies with low \HI{} demonstrate a perpendicular spin.

According to observational data, bright spiral galaxies predominantly exhibit co-directional spins with respect to the filaments, while elliptical galaxies tend to have a perpendicular spin direction \citep{2013MNRAS.428.1827T, 2013ApJ...775L..42T}. Other studies have found that galaxies located in denser environments predominantly display perpendicular spin orientations \citep{2007ApJ...671.1248L, 2010MNRAS.408..897J}. However, some authors did not identify any significant dependencies \citep{2016MNRAS.457..695P, 2019ApJ...876...52K, 2023MNRAS.522.4740K}.

Such differences in the results may stem from the challenges in determining the spin orientation of galaxies, identifying the filaments themselves, and the limited statistical sample sizes. 
There are two main difficulties in determining the direction of the normal to a galaxy's disk: significant measurement errors in the inclination of arbitrarily oriented galaxies and the ambiguity regarding which side of the galaxy is closest to the observer. 
Both of these issues can be effectively addressed by selecting a specific subset of galaxies. For edge-on galaxies, the inclination is, by definition, close to 90 degrees, with an accuracy better than 5 degrees. This significantly reduces the ambiguity in determining the direction of the disk's normal. Furthermore, the positional angle of the major axis can be determined with an accuracy better than 1 degree, allowing for a precise determination of the normal direction --- an accuracy that is typically unreachable for arbitrarily oriented galaxies.

In this paper, we aim to determine whether there is a preferred spin direction of disk galaxies relative to the filament axes of large-scale structures. For this analysis, we used the largest Edge-on Galaxy catalogue~\citep[EGIPS,][]{2022MNRAS.511.3063M} together with the filament catalogue~\citep{2014MNRAS.438.3465T}.

\section{Sample}

The uncertainty of the spin position can easily be circumvented by using edge-on galaxies. Currently, the EGIPS catalogue~\citep{2022MNRAS.511.3063M} represents the largest sample of edge-on galaxies, comprising 16,551 objects selected using an artificial neural network based on data from the second release of the Panoramic Survey Telescope and Rapid Response System (Pan-STARRS1). This catalogue includes a wide range of parameters, such as the position angle, which is essential to determine the spin. Completeness tests show that the EGIPS catalogue is 96\% complete for objects with a major semi-axis, $a_r>5.5^{\prime\prime}$.
Moreover, taking into account that the typical image quality in the Pan-STARRS1 survey is 1.1--$1.3^{\prime\prime}$~\citep{2020ApJS..251....6M}, the size- and axis-ratio estimates for the smallest EGIPS galaxies may suffer due to image blurring.
Thus, we excluded from consideration all edge-on galaxies with a major semi-axis $a_r<5^{\prime\prime}$.

One of the largest and most reliable filament catalogues~\citep{2014MNRAS.438.3465T} was created using the Sloan Digital Sky Survey (SDSS) data release~8 based on a sample of 499,340 galaxies with a redshift of $0.009 < z < 0.155$ in the cosmic microwave background (CMB) system.
It contains 15421 filaments.

We cross-identified galaxies from the EGIPS catalogue with the SDSS sample used to detect the filaments of the large-scale structure.
This process resulted in a sample of 2,831 edge-on galaxies, each associated with one of the 2,831 filaments.

It is important to note that each filament in the catalogue \citep{2014MNRAS.438.3465T} was constructed from an average of 32 galaxies. Given that the probability of a galaxy being edge-on is approximately 1\%, we would expect to find no more than one edge-on galaxy per filament, which aligns with our observations in this sample.

\section{Coordinate transformation}

The filament catalogue includes coordinates that describe the filament structure in the Cartesian coordinate system of SDSS ($\eta, \lambda$). In contrast, the positions of the galaxies are provided in the equatorial coordinate system, with corresponding position angles from the EGIPS catalogue also given in this system.To work with all this data, it is essential to convert all coordinates into a single reference frame.
For this purpose, we selected a Cartesian coordinate system where the $z$-axis points toward the north celestial pole, the $x$-axis is aligned with the vernal equinox, and the $x-y$ plane corresponds to the equatorial plane. This choice simplifies a analysis of the filaments and galaxies.

To convert filament coordinates, we use Equations~(\ref{eq:0}--\ref{eq:9}).
The coordinates of the filament points are first translated into the SDSS coordinate system ($\eta, \lambda$), according to \citet{2013MNRAS.428.1827T}:
\begin{align}
\alpha_{cen}     &= 185.0^{\circ} 
\label{eq:0}\\
\delta_{cen}    &= 32.5^{\circ}
\label{eq:1}\\
D       &= \sqrt{(x{^2} + y{^2} + z{^2})} \label{eq:2}\\
\lambda &= \arcsin(-x/D) \label{eq:3}\\
\eta    &= \arctan(z/y), \label{eq:4}
\end{align}
where D is the distance in Mpc\,h$^{-1}$. 

Then, to convert to the equatorial coordinates ($ \alpha,  \delta$) we use the formulas from the coordinate conversion software\footnote{\url{https://www.sdss4.org/dr17/algorithms/surveycoords/}}:

\begin{align}
\alpha &= \arctan \left( \frac{\sin(\lambda)}{\cos(\lambda)\cos(\eta + \delta_{cen})} \right) + \alpha_{cen} \label{eq:5}\\
\delta  &= \arcsin( \cos(\lambda) \sin(\eta + \delta_{cen}) ) \label{eq:6}
\end{align}

The Cartesian coordinate system ($x$, $y$, $z$), where the $z$-axis is directed to the north celestial pole, the $x$-axis is directed to the vernal equinox point, the $x-y$ plane lies in the equatorial plane:

\begin{align}
x &= D \cos(\delta) \cos(\alpha) \label{eq:7}\\
y &= D \cos(\delta) \sin(\alpha) \label{eq:8}\\
z &= D \sin(\delta) \label{eq:9}
\end{align}

Equations~(\ref{eq:7}--\ref{eq:9}) were used to convert the galaxy coordinates.

As the spin of an edge-on galaxy, we can use a position angle (PA) rotated by 90$^\circ$. To determine the spin vector in the Cartesian coordinate system, the rotation matrices were used, with the help of which the unit vector was rotated to a position corresponding to the spin of the galaxy under consideration. Initially, the unit vector has $\alpha = 0^\circ$, $\delta = 0^\circ$, $\mathrm{PA} = 90^\circ$, such a vector in the Cartesian coordinate system will have a very convenient representation $x = 0$, $y = 0$, $z = 1$. The coordinates of the galaxy spins in the Cartesian coordinate system were found using Equations~(\ref{eq:10}--\ref{eq:15}).

\begin{eqnarray}
s = \mathrm{PA}^{\circ} +90^\circ
\label{eq:10} 
\end{eqnarray}

\begin{eqnarray}
R_x(s) = \left(
\begin{array}{ccc}
1 & 0 & 0 \\
0 & \cos(-s) & -\sin(-s)\\
0 & \sin(-s) & \cos(-s)
\end{array}
\right)
 \label{eq:11}
 \end{eqnarray}
 
\begin{eqnarray}
R_y(\delta) = \left(
\begin{array}{ccc}
\cos(-\delta) & 0 & \sin(-\delta)\\                                  
0  & 1 & 0\\
-\sin(-\delta) & 0 & \cos(-\delta)
\end{array}
\right)
\label{eq:12}
\end{eqnarray}

\begin{eqnarray}
R_z(\alpha) = \left(
\begin{array}{ccc}
\cos(\alpha) & -\sin(\alpha) & 0\\
\sin(\alpha) & \cos(\alpha) & 0\\
0 & 0 & 1
\end{array}
\right)
\label{eq:13}
\end{eqnarray}

\begin{eqnarray}
n = \left(
\begin{array}{c}
0\\
0\\
1
\end{array}
\right)
\label{eq:14}
\end{eqnarray}

\begin{eqnarray}
n' = R_z R_y R_x n  
\label{eq:15}
\end{eqnarray}

Since filaments have a complex shape, we approximate it by a cubic spline without smoothing. 
We use the tangent at the filament point closest to our galaxy to determine the angle between the filament and the galaxy spin. 
The tangent was defined as the first derivative of the spline at this point.
The angle between the tangent to the filament and the spin of the galaxy is determined through the scalar product of the vectors.

\section{Galaxy-filament alignment}

We analyse the probability density of the angle, $\Theta$, between the galaxy spin and the filament axis of the large-scale distribution of galaxies.
The random distribution of angles between vectors in a three-dimensional space is uniform in $\cos\Theta$.
If there is no preferred spin direction relative to the filament axis, then there should be no trend in the distribution, and the linear regression slope should be equal to zero.
We should test the null hypothesis that the spins are randomly distributed with respect to the filaments, and, consequently, the slope of the distribution as a function of $\cos\Theta$ is zero.

\begin{figure}
\centering
\includegraphics[width=\linewidth]{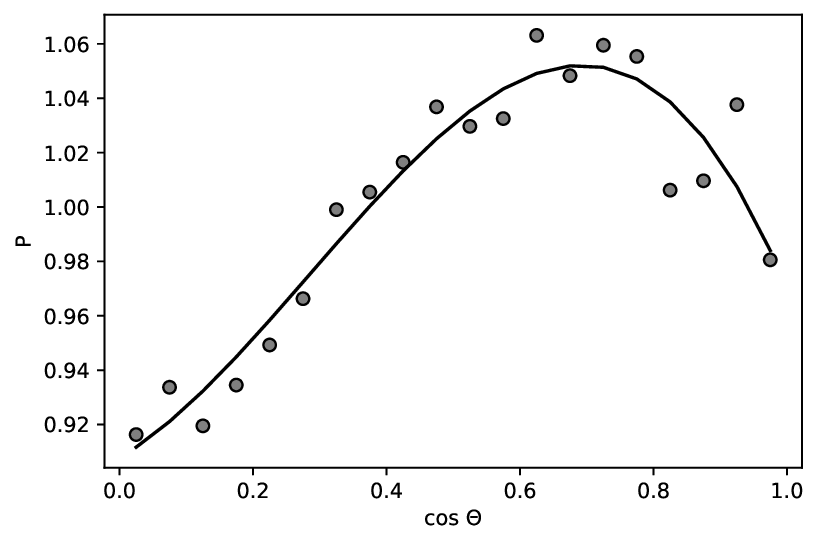}
\caption{Null hypothesis. 
The dependence of the probability density of the angle between the randomly oriented spin of the galaxies under study and the filament axes to which they belong. 
The solid black curve is the approximation of the points by a smoothing cubic spline.
The apparent correlation is a consequence of selection effects in creating the filaments that are predominantly oriented perpendicular to the line of sight, and in using the edge-on galaxies whose rotation axis is by definition oriented perpendicular to the line of sight.
This function describes the null hypothesis against which further analysis is carried out.
}
\label{fig:null-hypothesis}
\end{figure}

However, the observed filament distribution is systematically biased, with filaments preferentially oriented perpendicular to the line of sight. When combined with edge-on galaxies, whose spin direction is, by definition, also perpendicular to the line of sight, one would expect an artificial alignment even in the absence of any internal alignment between the galaxies and filaments.
To account for this effect, we generated a sample of $2.86\times10^6$ objects around the same 2861 filaments selected for analysis. The position angles of these test galaxies were randomly distributed across the sky. The probability density of the angle distribution was calculated in bins uniformly distributed along $\cos\Theta$.

Figure~\ref{fig:null-hypothesis} reproduces the probability density of the angles between the spins of randomly oriented galaxies and filament axes for the null hypothesis. It is evident that this probability density is quite complex and differs significantly from the naively expected constant value. The black curve represents a cubic spline fit to the distribution. To approximate the distribution, we used the Python function "UnivariateSpline()" with parameters k=3, s=10, and ext=3.

\begin{figure}
\includegraphics[width=\columnwidth]{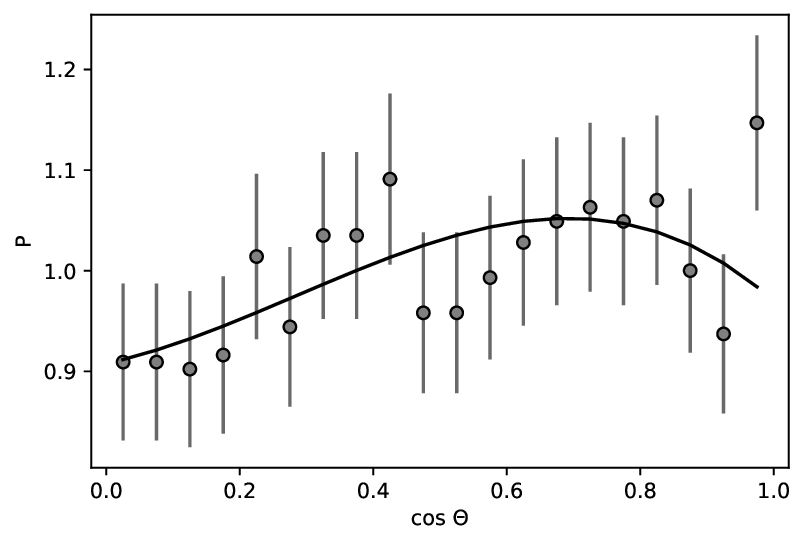}
\includegraphics[width=\columnwidth]{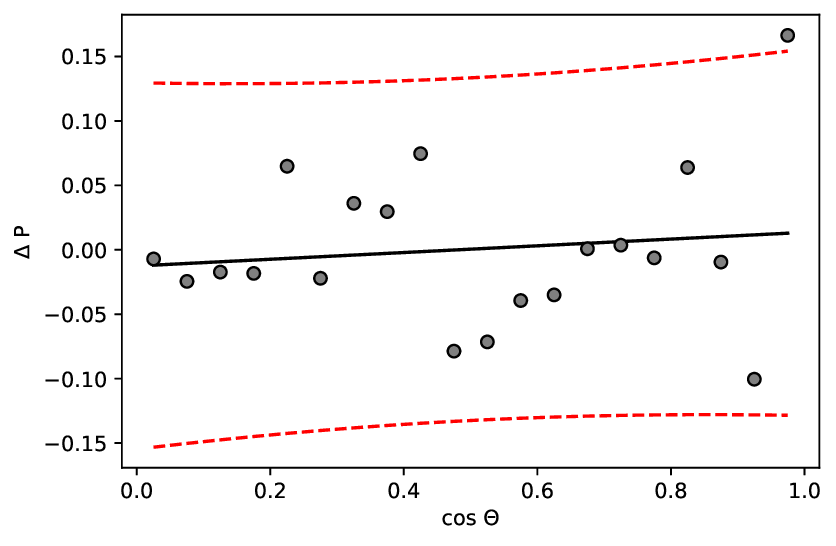}
\caption{
\textit{Upper panel}: the probability density, P, of the mutual orientation of the spins of galaxies and filaments in three-dimensional space. 
The black solid curve is the smoothed approximation by a cubic spline.
\textit{Bottom panel}: the probability density difference, $\Delta$ P, of mutual orientation between the observed spins of galaxies relative to the filaments in three-dimensional space and the null hypothesis (see Figure~\ref{fig:null-hypothesis}). 
The black solid line is the linear approximation. 
The red dashed lines show the 5\% confidence interval for an individual measurements. 
}
\label{fig:pdf_whole_sample}
\end{figure}

Figure~\ref{fig:pdf_whole_sample} (upper panel) shows the probability density as a function $\cos\Theta$ between the galaxy spins and the filament axes for our sample of 2861 edge-on galaxies.

It can be seen that, in general, it follows the probability density of randomly oriented galaxies (black solid curve in Figure~\ref{fig:null-hypothesis}). 
After correction for this selection effect caused by the non-uniformity of filament orientations, the deviations were described by a simple linear regression (see Figure~\ref{fig:pdf_whole_sample}, bottom panel). 
A statistically significant slope should indicate the existence of a correlation or anticorrelation between the galaxy spin and the filament axis, which in turn may testify the influence of large-scale structure on the rotation axis of galaxies. 
We do not find a statistically significant signal in the angle probability density after correction for the shape of the artificial signal.
The slope is $0.03 \pm 0.05$ (see Table~\ref{tab:regression}).

The absence of correlation in the galaxy sample we studied may be attributed to the fact that the initial angular momentum of protogalaxies is not conserved, leading to random spin directions of the galaxies. However, according to modern theory of evolution, early-type galaxies undergo mergers with galaxies of comparable sizes, which can affect the orientation of their rotation axes. In contrast, late-type galaxies may retain their initial rotation direction~\citep{2013MNRAS.428.1827T, 2013ApJ...775L..42T}. Furthermore, \citet{2012MNRAS.421L.137L} suggest that some early-type galaxies can dramatically change their spin orientation, with the new orientation becoming perpendicular to the initial one.

From this, we can conclude that by studying a sample that includes galaxies with diverse properties, we may not observe any dependencies in the alignment of their spins relative to large-scale structures, particularly filaments. Analysing a sample where one subset of galaxies has a preferred spin orientation relative to the filaments that is diametrically opposite to that of another subset risks yielding a uniform distribution, as illustrated in Figure~\ref{fig:pdf_whole_sample}.

Therefore, we will further analyse various subsamples selected by colour, distance from the filament, redshift, and other criteria. A more refined selection of galaxies may enable us to detect correlations, despite some degradation in statistics due to smaller sample sizes.

\subsection{Dependence on the colour index}

According to several studies (see Section~\ref{sec:Intro}), the orientation of galaxy spins may depend on various factors, including galaxy morphology.
This is not surprising, as modern studies suggest that galaxies of different morphological types are formed through distinct processes, and the environment plays a direct role in shaping the formation.
At this stage of the analysis, we examine whether the correlation depends on galaxy morphology.
The general galaxy population exhibits a bimodal distribution in the colour–absolute magnitude diagram. 
\citet{2004ApJ...608..752B} were the first to describe the individual areas of the diagram. 
The red sequence is populated by early-type galaxies, while the blue cloud is formed by spirals. 
A transitional region known as the 'green valley' lies between them, commonly interpreted as a zone of rapid evolution from late-type to early-type galaxies.
The edge-on EGIPS galaxies are well separated into the red sequence and the blue cloud by a colour index of $(g-i)_0 \approx 1$~\citep[see figures~8--10]{2022MNRAS.511.3063M}. 
The blue subsample, defined by $(g-i)_0 < 1$, includes 1,190 galaxies, while the red subsample, with $(g-i)_0>1$, contains 1,671 galaxies.
In both cases, the probability density distributions do not show statistically significant deviations from the null hypothesis of no correlation between galaxy spin orientations and filament axes.
The regression slopes are $-0.02 \pm 0.05$ for the blue subsample and $0.05 \pm 0.10$ for the red subsample.

\begin{figure}
\includegraphics[width=\columnwidth]{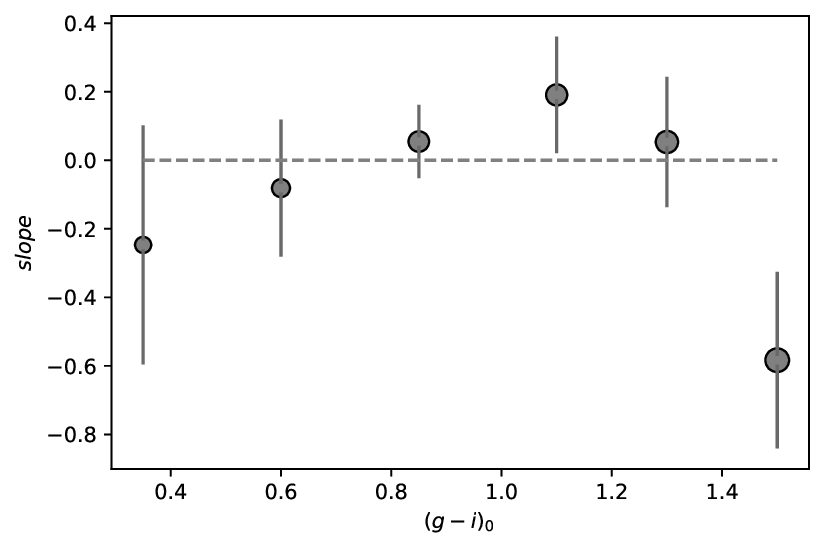}
\caption{
Dependence of the regression slope on the subsample colour. 
The dotted straight line shows the zero level. The size of the circle increases as the galaxies become brighter.
}
\label{fig:pvalue_gi}
\end{figure}

However, a more detailed division of the sample by colour index uncovered notable patterns of behaviour. Figure~\ref{fig:pvalue_gi} shows the regression slopes representing deviations from the null hypothesis for various subsamples, plotted as a function of the angle between the galaxy spin and filament axis. 
On the one hand, none of the individual subsamples differs significantly from the null hypothesis, except for the reddest subsample.
On the other hand, the subsamples exhibit regular behaviour that may suggest the presence of a correlation within a specific subset of galaxies.

\begin{figure}
\centering
\includegraphics[width=\columnwidth]{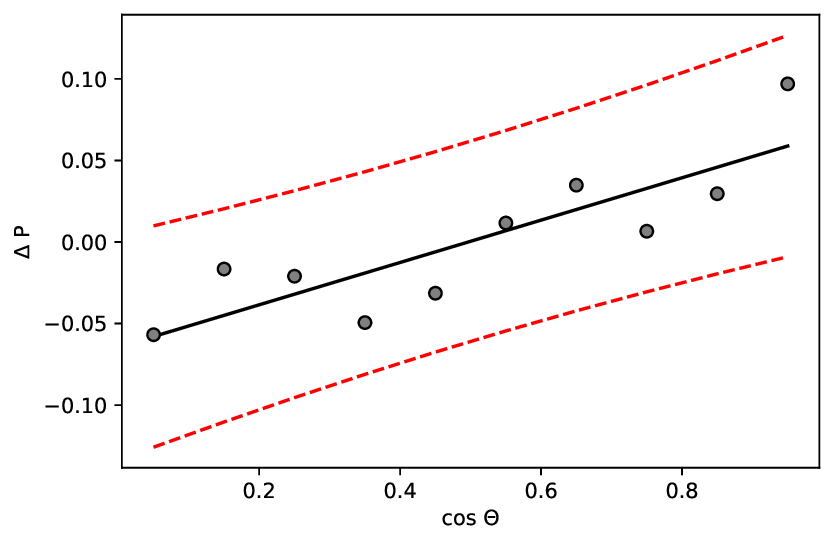}
\includegraphics[width=\columnwidth]{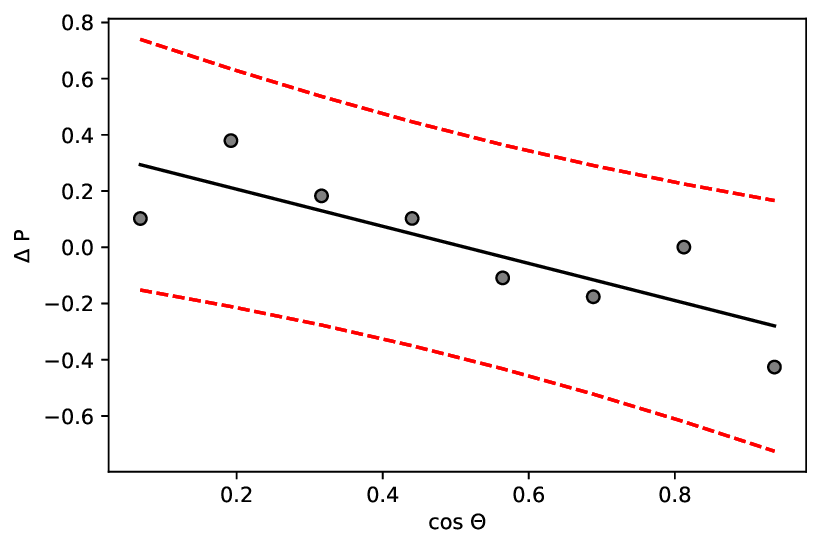}
\caption{
Deviation of the probability density from the null hypothesis of no correlation between galaxy spin and filament axis in three-dimensional space.
The black solid line is the linear regression. 
The red dotted lines show the 5\% confidence interval for an individual measurement.
The top panel shows the case of the subsample of 1890 edge-on galaxies with $0.7<(g-i)_0<1.3$.
The bottom panel shows the case of the subsample of 165 reddest galaxies with $(g-i)_0>1.4$.
}
\label{fig:density_cos_teta_color}
\end{figure}

The subsample of 1,890 edge-on galaxies with colours in the range $0.7<(g-i)_0<1.3$ demonstrates a tendency for their spins to align with the filament axes (see the top panel of Figure~\ref{fig:density_cos_teta_color}).
The regression slope is $0.13 \pm 0.03$ with the $p$-value of $0.002$.

The reddest subsample, with $(g-i)_0 > 1.4$, exhibits a negative regression slope of $-0.66 \pm 0.20$ with $p$-value of $0.01$, indicating a statistically significant trend toward a perpendicular orientation of galaxy spins relative to filament axes (see the bottom panel of Figure~\ref{fig:density_cos_teta_color}).
The lower statistical significance of this result may be attributed to the small sample size of only 165 galaxies.

We expected galaxies with $(g-i)_0 > 1$ to have their spins oriented perpendicular to the filaments. This expectation is based on the idea that early-type galaxies form primarily through major mergers occurring along filaments. During these events, the orbital angular momentum of the merging systems is converted into intrinsic spin, resulting in a spin orientation that is typically perpendicular to the filament direction. However, we observe this trend only among galaxies with $(g-i)_0 > 1.4$. 
This observed result can be explained by the insufficient mass of early-type galaxies with  $(g-i)_0 > 1.4$. According to \citet{2012MNRAS.421L.137L}, only galaxies with masses exceeding $(8\pm2)\times10^{12}$~M$_\odot$ exhibit perpendicular spin. Galaxies with $(g-i)_0 > 1.4$ are the brightest in our sample, as indicated by the circle size in Figure~\ref{fig:pvalue_gi}. Their red colours further suggest that they are also the most massive.

\subsection{Dependence on different parameters} 

To enhance the alignment signal, we investigate how various galaxy properties, such as absolute magnitude in the $r$-band, distance from the filament, and redshift, influence the correlation between galaxy spin orientations and filament axes.

\citet{2013MNRAS.428.1827T} found that the correlation is stronger for bright galaxies.
The strongest parallel alignment is observed among galaxies with $M_r < -20.7$. 
In our case, we find a good correlation for galaxies brighter $-19$.

\begin{figure}
\includegraphics[width=\columnwidth]{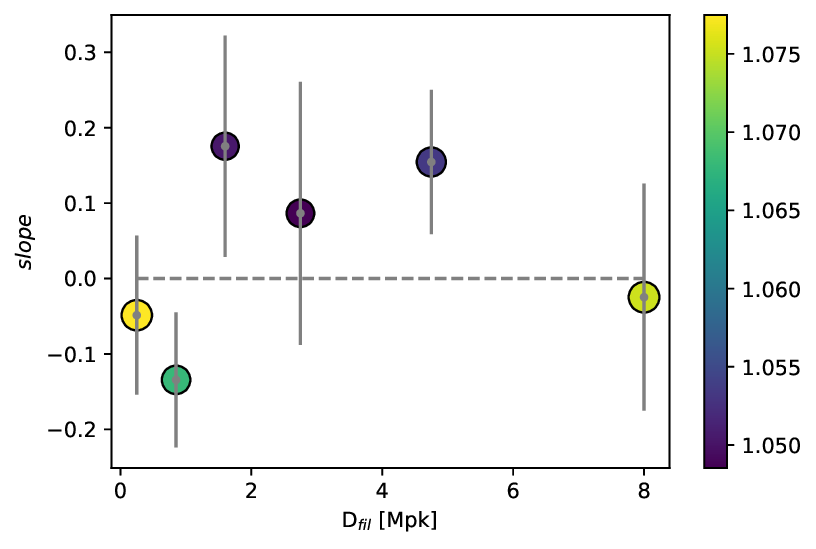}
\caption{
Mean regression slope as function of the distances between filament and galaxy. The colour shows the $g-i$ colour index of galaxies included in the indicated range.
}
\label{fig:slope_distance_to_filament}
\end{figure}

Figure~\ref{fig:slope_distance_to_filament} suggests that, on average, galaxies tend to align their spin with the filament axis with increasing distance from the filament axis.
This correlation can be observed at sufficiently large distances up to 6~Mpc, but due to the large scatter of estimates, no unambiguous conclusion can be drawn and it remains only a hint of the effect.
However, it should be noted that such large-scale correlations on scales up to 6~Mpc have been found previously~\citep{2019ApJ...884..104L, 2022ApJ...935...71K}. 

Also Figure~\ref{fig:slope_distance_to_filament} demonstrates that the reddest galaxies are located near the filament axis, while the blue galaxies prefer to settle on larger distances. However, this effect is very subtle.
\citet{2017MNRAS.465.3784B} shows a dependence of galaxy type on distance to the filament, with different types of galaxies having different spin positions relative to the filament. They find that blue galaxies are located farther from the filament axis. \citet{2017A&A...600L...6K} also found this effect.

\begin{figure}
\includegraphics[width=\columnwidth]{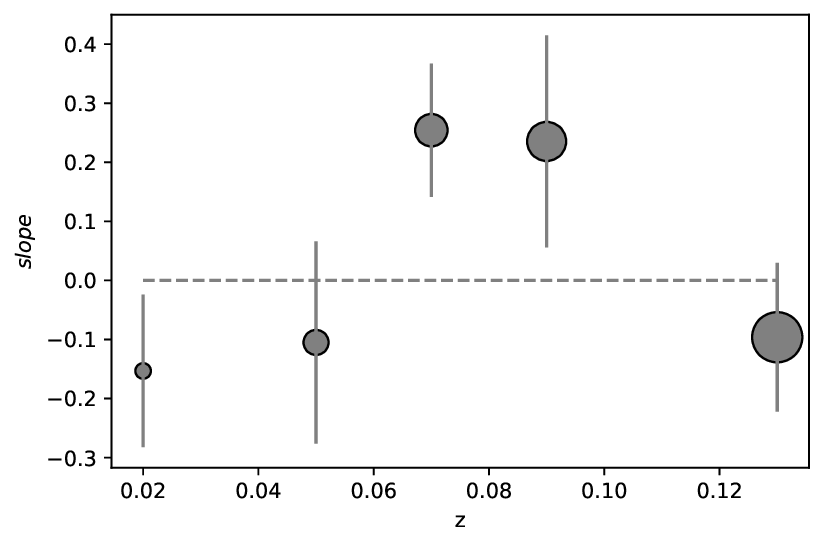}
\caption{
Mean regression slope as function of redshift. 
The circle size is proportional to the mean galaxy absolute magnitude in the subsample
}
\label{fig:slope_redshift}
\end{figure}

According to Figure~\ref{fig:slope_redshift}, galaxies within the redshift range $0.06<z<0.10$ exhibit aligned spin orientations, whereas the correlation disappears for smaller and higher redshifts.
This can be explained by boundary effects.
The volume of the nearby Universe is rather small, and filaments close to the survey boundary are affected by the survey edge effect. This might affect the alignment for nearby bins. 
For far away bins, the statistics is probably not good enough, and the galaxy density is lower, which is also affecting the filament detection. In two intermediate redshift bins, we have sufficient statistics and the most reliable filaments.

\begin{figure}
\includegraphics[width=\columnwidth]{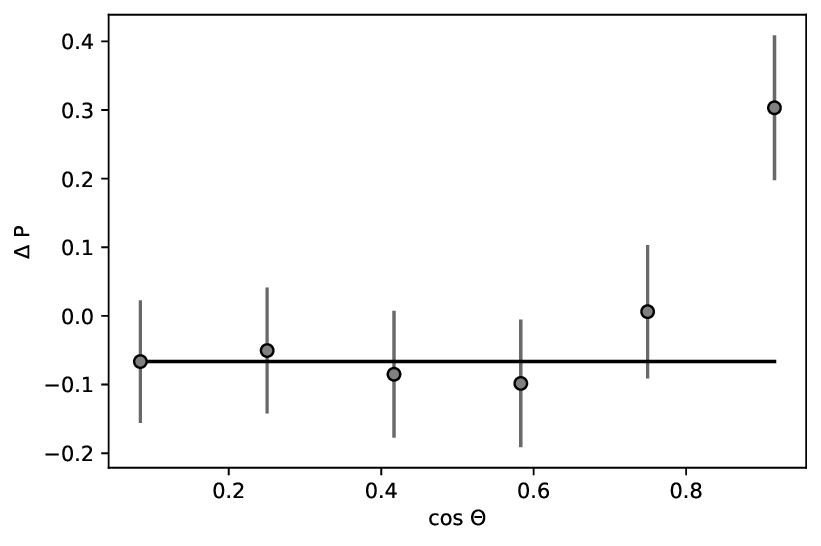}
\caption{
Deviation of the probability density from the null hypothesis of no correlation between galaxy spins and filament axes in three-dimensional space. 
The distribution is based on a sample of 554 edge-on galaxies width $0.7<(g-i)_0<1.3$, $M_r<-19$, $0.06<z<0.10$. 
}
\label{fig:density_many_param}
\end{figure}

Combining together the conditions associated with the strongest correlations $0.7<(g-i)_0<1.3$ \& $M_r<-19$ \& $0.06<z<0.10$, we obtain a subsample of 554 edge-on galaxies.
Figure~\ref{fig:density_many_param} illustrates the existence of a statistically significant predominance of galaxies with spin codirected with filament over the background of randomly oriented galaxies.
The probability density at the last point with $\cos\Theta\sim1$ is 3.7 $\sigma$ above the median average.

\subsection{Correlation between galaxy spins and filament axes}

The results of the linear regression analysis are summarised in Table~\ref{tab:regression}. 
The first column, condition, indicates the subsample selection criteria; 
the slope characterises the deviation from the null hypothesis, namely deviation from the random orientation of spins with respect to the filament axis; 
the third column, std err, gives the standard error of the slope; 
the fourth column, N, stays for the number of galaxies in the sample; 
the $p$-value is the probability of the given result under the assumption that the slope equals zero (the most common threshold is $p$-value $< 0.05$).

\begin{table}[hbt!]
\begin{threeparttable}
\caption{Correlation between galaxy spins and filament axes}
\label{tab:regression}
\begin{tabular}{lllll}
\toprule
\headrow Condition & slope  & std err & N & $p$-value\\
\midrule
all & 0.03 & 0.05 & 2861 & 0.6\\ 
\midrule
$(g-i)_0$ < 1 & -0.02 & 0.05 &1190 & 0.7\\
\midrule
$(g-i)_0$ > 1 & 0.05 & 0.10 &1671 & 0.6\\
\midrule
 0.7 < $(g-i)_0$ < 1.3 & 0.13 & 0.03 & 1890 & 0.002\\
 \midrule
 $(g-i)_0$ > 1.4 & -0.7 & 0.2 & 165 & 0.01\\
\bottomrule
\end{tabular}
\begin{tablenotes}[hang]
\item[Condition] is the selection criterion;
\item[slope] is the slope of the probability density distribution;
\item[std err] is the standard error of the slope;
\item[N] is the number of galaxies in the sample.
\end{tablenotes}
\end{threeparttable}
\end{table}

\section{Conclusion}

In this paper, we examine the three-dimensional orientation of the edge-on galaxy spin axes relative to the filaments of the large-scale structure of the Universe. 
Edge-on galaxies allow us to determine unambiguously and with great precision the directions of their spin in space.
The main limitation of using edge-on galaxies is their relatively small number, as they represent only about 1\% of all disk galaxies.
The study is based on the largest to date sample of 16,551 edge-on galaxies selected in the Pan-STARRS1 survey~\citep[EGIPS,][]{2022MNRAS.511.3063M} and the most reliable catalogue of 15421 filaments~\citep{2014MNRAS.438.3465T} created using the SDSS DR8 redshift survey.
After cross-identification of these catalogues, our final sample of edge-on galaxies comprises 2861 objects.

After accounting for observational effects, we look for deviations from the uniform distribution of the probability density of the $\cos\Theta$ between spin and filament axis. 
We identify statistically significant correlations in the two galaxy subsamples above 3 $\sigma$ significance level.

The first is the subsample of 1890 galaxies with $0.7<(g-i)_0<1.3$.
Their probability density increases with angle between spins and filaments as $(0.13 \pm 0.03) \cos\Theta$.
This implies that the spins of these galaxies tend to align with the filaments at a confidence level of 0.2\% corresponding to 4.3 $\sigma$.
This result agrees well with the works by \citet{2007ApJ...655L...5A, 2012MNRAS.427.3320C, 2012MNRAS.421L.137L, 2013ApJ...762...72T}. 
According to theoretical models, low-mass galaxies are formed through the winding of flows embedded in misaligned cosmic walls.
As a result, they acquire spin orientations parallel to the axes of the filaments that form at the intersection of these walls. 

The second subsample consists of only 165 reddest galaxies with $(g-i)_0 > 1.4$.
Its probability density demonstrates a negative correlation $\sim (-0.7 \pm 0.2) \cos\Theta$ with $p$-value of 0.01 or 3.5 $\sigma$.
Thus, they appear to be mostly oriented perpendicular to the filaments.
The described sample consists of early-type lenticular galaxies. 
It is assumed that early-type galaxies are formed through major merger occurring along filaments.
Consequently, they acquire spin orientations perpendicular to the merger direction as their orbital angular momentum is transformed into intrinsic spin. 
According to \citet{2012MNRAS.421L.137L}, perpendicularly oriented spin should be in galaxies with masses higher than $(8 \pm 2) \times 10^{12}$~$M_\odot$. 

The probability density of the sample of galaxies with $0.7<(g-i)_0<1.3$ \& $M_r<-19$ \& $0.06<z<0.10$ demonstrates the sharp excess of galaxies with spins almost parallel to the filaments.
This point deviates by $3.7\sigma$ from the population of other randomly oriented galaxies.
This result can be interpreted as the presence of a population of galaxies that have the spin co-directed to the filament axis.

There is an indication of spin-filament alignment on scales up to 6~Mpc from the filament, although this trend is very weak due to the substantial scatter in the data.
Nevertheless, such unexpected large-scale correlations between kinematics of giant galaxies and the motion of surrounding galaxies within 6~Mpc have been reported by \citet{2019ApJ...884..104L, 2022ApJ...935...71K}.

\section{Acknowledgements}
The authors thank Alexander Marchuk for fruitful discussions.
AVA and DIM acknowledge the support of the Russian Science Foundation grant~\textnumero~24--72--10084. 
ET acknowledges funding from the HTM (grant TK202), ETAg (grant PRG1006) and the EU Horizon Europe (EXCOSM, grant No. 101159513).
We acknowledge the usage of the HyperLeda\footnote{\url{http://leda.univ-lyon1.fr}} database~\citep{2014A&A...570A..13M}.

\printendnotes

\bibliography{main.bib}

\begin{thebibliography}{}
\expandafter\ifx\csname natexlab\endcsname\relax\def\natexlab#1{#1}\fi

\bibitem[{{Arag{\'o}n-Calvo} {et~al.}(2007){Arag{\'o}n-Calvo}, {van de Weygaert}, {Jones}, \& {van der Hulst}}]{2007ApJ...655L...5A}
{Arag{\'o}n-Calvo}, M.~A., {van de Weygaert}, R., {Jones}, B. J.~T., \& {van der Hulst}, J.~M. 2007, \apjl, 655, L5

\bibitem[{{Bell} {et~al.}(2004){Bell}, {Wolf}, {Meisenheimer}, {Rix}, {Borch}, {Dye}, {Kleinheinrich}, {Wisotzki}, \& {McIntosh}}]{2004ApJ...608..752B}
{Bell}, E.~F., {Wolf}, C., {Meisenheimer}, K., {et~al.} 2004, \apj, 608, 752

\bibitem[{{Bizyaev} {et~al.}(2017){Bizyaev}, {Kautsch}, {Sotnikova}, {Reshetnikov}, \& {Mosenkov}}]{2017MNRAS.465.3784B}
{Bizyaev}, D.~V., {Kautsch}, S.~J., {Sotnikova}, N.~Y., {Reshetnikov}, V.~P., \& {Mosenkov}, A.~V. 2017, \mnras, 465, 3784

\bibitem[{{Codis} {et~al.}(2012){Codis}, {Pichon}, {Devriendt}, {Slyz}, {Pogosyan}, {Dubois}, \& {Sousbie}}]{2012MNRAS.427.3320C}
{Codis}, S., {Pichon}, C., {Devriendt}, J., {et~al.} 2012, \mnras, 427, 3320

\bibitem[{{Dubinski}(1992)}]{1992ApJ...401..441D}
{Dubinski}, J. 1992, \apj, 401, 441

\bibitem[{{Efstathiou} \& {Jones}(1979)}]{1979MNRAS.186..133E}
{Efstathiou}, G., \& {Jones}, B.~J.~T. 1979, \mnras, 186, 133

\bibitem[{{Ganeshaiah Veena} {et~al.}(2019){Ganeshaiah Veena}, {Cautun}, {Tempel}, {van de Weygaert}, \& {Frenk}}]{2019MNRAS.487.1607G}
{Ganeshaiah Veena}, P., {Cautun}, M., {Tempel}, E., {van de Weygaert}, R., \& {Frenk}, C.~S. 2019, \mnras, 487, 1607

\bibitem[{{Hahn} {et~al.}(2010){Hahn}, {Teyssier}, \& {Carollo}}]{2010MNRAS.405..274H}
{Hahn}, O., {Teyssier}, R., \& {Carollo}, C.~M. 2010, \mnras, 405, 274

\bibitem[{{Jones} {et~al.}(2010){Jones}, {van de Weygaert}, \& {Arag{\'o}n-Calvo}}]{2010MNRAS.408..897J}
{Jones}, B. J.~T., {van de Weygaert}, R., \& {Arag{\'o}n-Calvo}, M.~A. 2010, \mnras, 408, 897

\bibitem[{{Karachentsev} \& {Zozulia}(2023)}]{2023MNRAS.522.4740K}
{Karachentsev}, I.~D., \& {Zozulia}, V.~D. 2023, \mnras, 522, 4740

\bibitem[{{Kim} {et~al.}(2022){Kim}, {Smith}, \& {Shin}}]{2022ApJ...935...71K}
{Kim}, Y., {Smith}, R., \& {Shin}, J. 2022, \apj, 935, 71

\bibitem[{{Kraljic} {et~al.}(2019){Kraljic}, {Pichon}, {Dubois}, {Codis}, {Cadiou}, {Devriendt}, {Musso}, {Welker}, {Arnouts}, {Hwang}, {Laigle}, {Peirani}, {Slyz}, {Treyer}, \& {Vibert}}]{2019MNRAS.483.3227K}
{Kraljic}, K., {Pichon}, C., {Dubois}, Y., {et~al.} 2019, \mnras, 483, 3227

\bibitem[{{Krolewski} {et~al.}(2019){Krolewski}, {Ho}, {Chen}, {Chan}, {Tenneti}, {Bizyaev}, \& {Kraljic}}]{2019ApJ...876...52K}
{Krolewski}, A., {Ho}, S., {Chen}, Y.-C., {et~al.} 2019, \apj, 876, 52

\bibitem[{{Kuutma} {et~al.}(2017){Kuutma}, {Tamm}, \& {Tempel}}]{2017A&A...600L...6K}
{Kuutma}, T., {Tamm}, A., \& {Tempel}, E. 2017, \aap, 600, L6

\bibitem[{{Lee} \& {Erdogdu}(2007)}]{2007ApJ...671.1248L}
{Lee}, J., \& {Erdogdu}, P. 2007, \apj, 671, 1248

\bibitem[{{Lee} \& {Pen}(2000)}]{2000ApJ...532L...5L}
{Lee}, J., \& {Pen}, U.-L. 2000, \apjl, 532, L5

\bibitem[{{Lee} {et~al.}(2019){Lee}, {Pak}, {Song}, {Lee}, {Kim}, \& {Jeong}}]{2019ApJ...884..104L}
{Lee}, J.~H., {Pak}, M., {Song}, H., {et~al.} 2019, \apj, 884, 104

\bibitem[{{Libeskind} {et~al.}(2013){Libeskind}, {Hoffman}, {Forero-Romero}, {Gottl{\"o}ber}, {Knebe}, {Steinmetz}, \& {Klypin}}]{2013MNRAS.428.2489L}
{Libeskind}, N.~I., {Hoffman}, Y., {Forero-Romero}, J., {et~al.} 2013, \mnras, 428, 2489

\bibitem[{{Libeskind} {et~al.}(2012){Libeskind}, {Hoffman}, {Knebe}, {Steinmetz}, {Gottl{\"o}ber}, {Metuki}, \& {Yepes}}]{2012MNRAS.421L.137L}
{Libeskind}, N.~I., {Hoffman}, Y., {Knebe}, A., {et~al.} 2012, \mnras, 421, L137

\bibitem[{{Magnier} {et~al.}(2020){Magnier}, {Schlafly}, {Finkbeiner}, {Tonry}, {Goldman}, {R{\"o}ser}, {Schilbach}, {Casertano}, {Chambers}, {Flewelling}, {Huber}, {Price}, {Sweeney}, {Waters}, {Denneau}, {Draper}, {Hodapp}, {Jedicke}, {Kaiser}, {Kudritzki}, {Metcalfe}, {Stubbs}, \& {Wainscoat}}]{2020ApJS..251....6M}
{Magnier}, E.~A., {Schlafly}, E.~F., {Finkbeiner}, D.~P., {et~al.} 2020, \apjs, 251, 6

\bibitem[{{Makarov} {et~al.}(2014){Makarov}, {Prugniel}, {Terekhova}, {Courtois}, \& {Vauglin}}]{2014A&A...570A..13M}
{Makarov}, D., {Prugniel}, P., {Terekhova}, N., {Courtois}, H., \& {Vauglin}, I. 2014, \aap, 570, A13

\bibitem[{{Makarov} {et~al.}(2022){Makarov}, {Savchenko}, {Mosenkov}, {Bizyaev}, {Reshetnikov}, {Antipova}, {Tikhonenko}, {Usachev}, {Borisov}, {Makarova}, {Kautsch}, {Marchuk}, \& {Rubtsov}}]{2022MNRAS.511.3063M}
{Makarov}, D., {Savchenko}, S., {Mosenkov}, A., {et~al.} 2022, \mnras, 511, 3063

\bibitem[{{Pahwa} {et~al.}(2016){Pahwa}, {Libeskind}, {Tempel}, {Hoffman}, {Tully}, {Courtois}, {Gottl{\"o}ber}, {Steinmetz}, \& {Sorce}}]{2016MNRAS.457..695P}
{Pahwa}, I., {Libeskind}, N.~I., {Tempel}, E., {et~al.} 2016, \mnras, 457, 695

\bibitem[{{Peebles}(1969)}]{1969ApJ...155..393P}
{Peebles}, P.~J.~E. 1969, \apj, 155, 393

\bibitem[{{Porciani} {et~al.}(2002){Porciani}, {Dekel}, \& {Hoffman}}]{2002MNRAS.332..325P}
{Porciani}, C., {Dekel}, A., \& {Hoffman}, Y. 2002, \mnras, 332, 325

\bibitem[{{Tempel} \& {Libeskind}(2013)}]{2013ApJ...775L..42T}
{Tempel}, E., \& {Libeskind}, N.~I. 2013, \apjl, 775, L42

\bibitem[{{Tempel} {et~al.}(2014){Tempel}, {Stoica}, {Mart{\'\i}nez}, {Liivam{\"a}gi}, {Castellan}, \& {Saar}}]{2014MNRAS.438.3465T}
{Tempel}, E., {Stoica}, R.~S., {Mart{\'\i}nez}, V.~J., {et~al.} 2014, \mnras, 438, 3465

\bibitem[{{Tempel} {et~al.}(2013){Tempel}, {Stoica}, \& {Saar}}]{2013MNRAS.428.1827T}
{Tempel}, E., {Stoica}, R.~S., \& {Saar}, E. 2013, \mnras, 428, 1827

\bibitem[{{Trowland} {et~al.}(2013){Trowland}, {Lewis}, \& {Bland-Hawthorn}}]{2013ApJ...762...72T}
{Trowland}, H.~E., {Lewis}, G.~F., \& {Bland-Hawthorn}, J. 2013, \apj, 762, 72

\bibitem[{{Wang} {et~al.}(2018){Wang}, {Guo}, {Kang}, \& {Libeskind}}]{2018ApJ...866..138W}
{Wang}, P., {Guo}, Q., {Kang}, X., \& {Libeskind}, N.~I. 2018, \apj, 866, 138

\bibitem[{{White}(1984)}]{1984ApJ...286...38W}
{White}, S.~D.~M. 1984, \apj, 286, 38

\end{thebibliography}

\appendix
\end{document}